\documentclass{iau}

\usepackage{amsmath}
\usepackage{graphicx}
\usepackage{multirow}

\begin{document}

\lefttitle{Allende Prieto}
\righttitle{Automated Methods}

\jnlPage{1}{7}
\jnlDoiYr{2025}
\doival{10.1017/xxxxx}
\volno{395}
\pubYr{2025}
\journaltitle{Stellar populations in the Milky Way and beyond}

\aopheadtitle{Proceedings of the IAU Symposium}
\editors{J. Mel\'endez,  C. Chiappini, R. Schiavon \& M. Trevisan, eds.}

\title{Automated Methods for Abundance Determination}

\author{Carlos Allende Prieto}
\affiliation{Instituto de Astrof\'{\i}sica de Canarias}

\begin{abstract}

As the multiplexing power of spectroscopic instruments increases, so 
does the need for automated analysis.  
In practice, the bottleneck for speed is the 
calculation of model spectra to evaluate the likelihood of candidate 
parameters. This presentation gives an overview of the steps required 
for automating spectroscopic analyses,  focusing on the speedups 
achievable by precomputing regular grids of synthetic spectra for 
on-the-fly interpolation, and a new 
technique based on precomputed irregular grids 
 capable of tackling problems with much higher 
dimensionality, as in the case when we are interested in deriving the 
abundances of multiple elements. Accuracy, ease of use and portability 
will be discussed.

\end{abstract}


\maketitle

\section{Introduction}

Stellar spectroscopy is traditionally a field in which observations
are made for one target at a time, especially for high-resolution instruments.
Data handling used to be a slow and demanding job up to the 1980s, 
when photographic plates
were the norm. While the advent of digital detectors and CCDs in particular facilitated
data reduction, spectroscopists performed data reduction and analysis
with software designed for interactive use.
This all changed at the turn of the century with the development of
multi-object spectrographs (see, e.g., \cite{1997SPIE.2871..145T,1999AAS...195.8701U}, in most cases taking advantage of optical fibers. 

We can get a glimpse at the importance of streamlining the analyses of
stellar spectra from a pioneering paper by \citet{1987PASP...99..335C},
including our
Beatrice Barbuy, using the Lick 3.1m telescope and the Robinson-Wampler image-dissector 
scanner to measure carbon and nitrogen abundances in 83 stars. This number of stars was already quite significant, 
considering the analysis of a single spectrum can take many hours.

\begin{figure}
  \includegraphics[scale=0.7,angle=0]{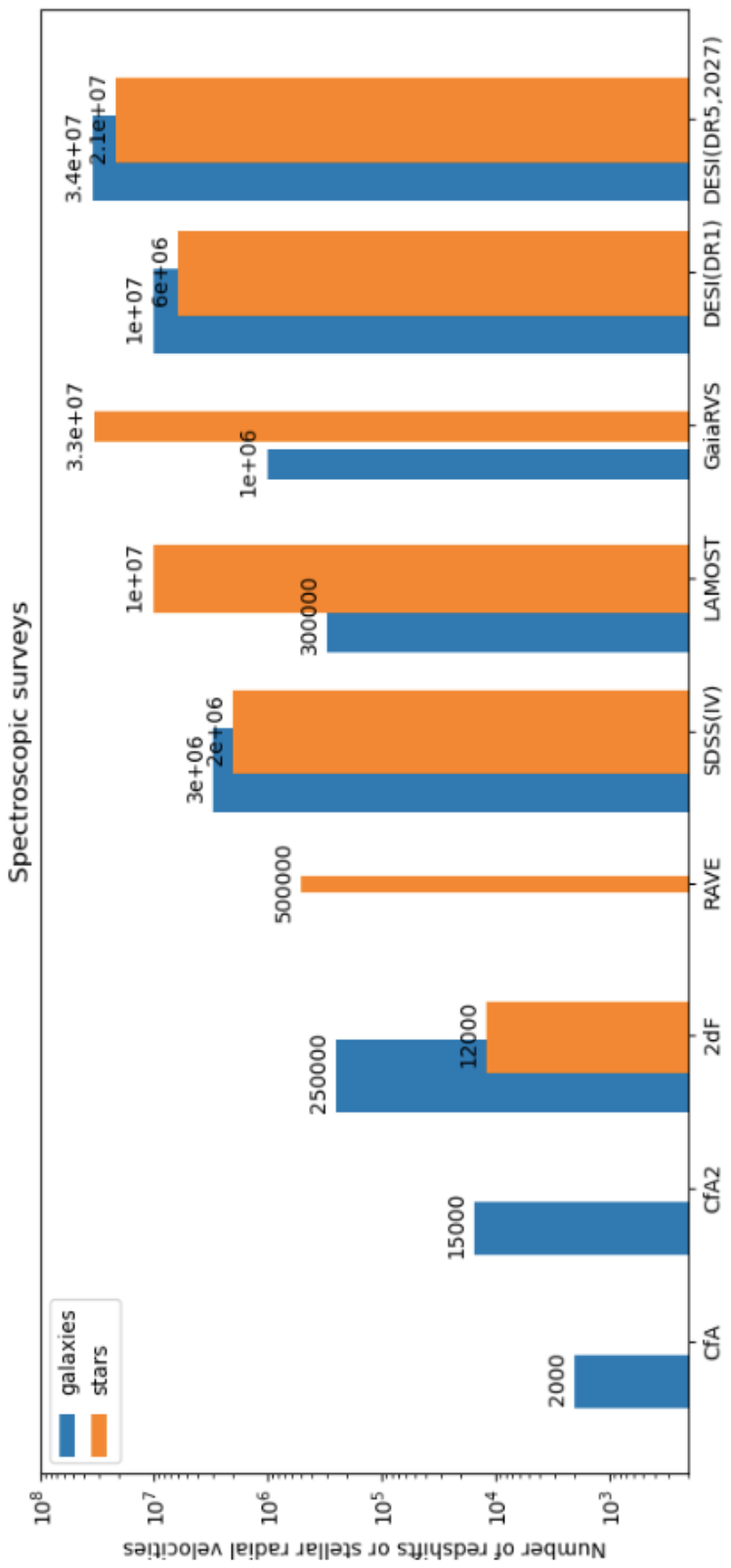}
  \caption{Evolution of the number of targets in spectroscopic surveys of stars and/or galaxies. The width of the bars reflects the survey's depth, with the Gaia RVS reaching $V\sim 16$ while DESI includes stars down to $\sim$ 19.}
  \label{f1}
\end{figure}

Automating the data reduction and analysis offers many advantages.  
First, it makes the results far more reproducible. Second, provided the 
software is written by expert hands, knowledgeable about the instrument, 
it brings optimal or nearly optimal results, independent from the user. 
Third, it makes data processing
much faster, demanding less human time. Given the continuous increase in 
field-of-view and multiplexing capability of spectrographs, with modern
surveys gathering data for $\sim 10^7$ targets over a few years (see Fig. 
\ref{f1}), 
automation is actually the only way we can cope with the data!

This article is focused on data analysis, taken as the the data processing 
that takes place after the spectra have been {\it reduced}, meaning they have
already been extracted (collapsed and background subtracted),  as well as 
wavelength- and flux-calibrated. The analysis involves the use of physical
models to infer information about the stellar source, and perhaps the 
intervening material.  

Classical optimization techniques have been the workhorse for automated spectral analysis.  The model parameters are those corresponding to the atmosphere of the star (effective temperature, $T_{\rm eff}$, surface gravity $\log g$, and
overall metal abundance [Fe/H]), the abundances of individual elements, and 
a few other parameters such 
as the line-of-sight velocity or the projected rotational velocity. The 
parameters are derived by optimization, searching for the combination that
gives the best agreement with the observed spectrum, sometimes supplemented with
photometric or astrometric data.

In the classical methodology, the information from spectra about the strength
of stellar absorption lines is condensed in {\it equivalent widths}, the
lines' area under the continuum, and the parameters are optimized one at a time.
Nowadays, $\chi-$square minimization is usually preferred, since it allows
using overlapping lines and excluding parts of line profiles affected 
by modeling imperfections, and the optimization 
is performed over many or all of the parameters simultaneously. 
The speed is typically limited by the time it takes to evaluate the $\chi-$square,
 which tends to be dominated by the solution of the radiative transfer equation 
 to compute a synthetic spectrum.

In recent years that machine learning is in fashion, it is common to apply, 
for example,  
convolutional neural networks to take spectra as input and return the atmospheric parameters. 
The networks need of course to be trained with the results from previous analyses, 
which are bound to be only available for a sample more limited 
than the target observations. This means
that rare stars may be under-represented in the 
training sample. There is also a risk that networks learn from built-in 
relationships in the training data, and  that those relationships no longer
apply in a larger or more distant sample.

\section{Analysis of stellar spectra}

Although there are references showing that the Romans were aware of 
the ability of a prism to spread light in colors, the introduction of the 
word {\it spectrum} 
is associated with Newton, and stellar spectroscopy 
could be considered born after the work by \cite{Wollaston1802}
and \cite{Fraunhofer1817}  with
solar light. The connection between spectral lines and chemical elements was set
of solid grounds by \cite{KB1860}. 

The analysis of stellar spectra in a modern physics-based setting 
starts by taking an informed guess at
the atmospheric parameters ($T_{\rm eff}$, $\log g$ and [Fe/H]), 
procuring a model atmosphere for that set
of parameters, computing a synthetic spectrum for that model,
and confronting that spectrum with observations, iterating  until
no further agreement can be found. Once the atmospheric parameters
are settled, the  analysis moves on to derive the abundances of
individual elements, which can be few or many depending on the 
quality of the data (mostly resolving power and signal-to-noise)
and the nature of the star. 

This process can be coded for a computer and therefore fully automated.
Over the next sections we will tackle, from a practical perspective, 
how to do this with existing models and tools. The scope of this essay is fairly limited,
since rather than producing a global overview of everything out there,
the focus is placed on how the author suggests to proceed. 
There is no doubt other authors will take different different paths.
A self-imposed restriction is to limit the discussion to cool
stars, mainly AFGKM, since those are the most abundant and interesting
for applications such as 
galactic chemical evolution, chemical tagging, or understanding 
the relationships between the chemistry of stars and the exoplanets
they host.

\section{Model atmospheres}

The recent review by \cite{2024arXiv240903329P} gives a comprehensive list of
stellar model atmosphere codes, as well as diagnostics software. For cool stars
the models computed with Kurucz's, MARCS and Phoenix codes are the most heavily used. The code-bases of MARCS or Phoenix are not public. However, models can
always be taken from existing grids.
One the possible choices in that case is to interpolate the model atmospheres,
which are basically tables describing the run of physical properties with
depth. An alternative, favored by \cite{2013MNRAS.430.3285M}, 
is to synthesize the
spectra for the available models, and then interpolate in them instead of 
in the model atmospheres. 
Model atmospheres can be computed afresh with open-source codes such as
Kurucz's \citep{1979ApJS...40....1K, 2005MSAIS...8...14K} or 
TLUSTY \citep{1995ApJ...439..875H, 2017arXiv170601859H}.

 The latest and largest grid of Kurucz models 
I am aware of is that by \cite{2012AJ....144..120M}, 
which includes a similar grid of MARCS models \cite{2008A&A...486..951G}. 
These grids were initially designed to satisfy the needs of the 
APOGEE survey \citep{2017AJ....154...94M} and are available from 
the pages of the \href{https://data.sdss.org/sas/dr17/apogee/spectro/speclib/atmos/}{Sloan Digital Sky Survey}
\citep{2022ApJS..259...35A}. These Kurucz models are also available from the 
\href{https://research.iac.es/proyecto/ATLAS-APOGEE/}{IAC}. Complementary sets 
of MARCS models can be downloaded from the \href{https://marcs.astro.uu.se}{MARCS website} and older Kurucz models, with a wider range in [Fe/H] and $T_{\rm eff}$ 
can be found at \href{http://kurucz.harvard.edu}{Kurucz's website}. Among the 
various choices of Phoenix \citep{1997ApJ...483..390H} grids, the most popular is described 
by \cite{2013A&A...553A...6H} and available from \href{https://www.astro.uni-jena.de/Users/theory/for2285-phoenix/grid.php}{Jena}.

Thomas Masseron has written a \href{https://marcs.astro.uu.se/software.html}{FORTRAN piece of code} for interpolating linearly 
MARCS model atmospheres. A similar 
tool for Kurucz models in IDL is available from \href{https://www.as.utexas.edu/~hebe/stools}{the University of Texas at Austin}, and works with the open-source GDL package. More recently, \cite{2023A&A...675A.191W} have trained a \href{https://github.com/cwestend/iNNterpol}{neural network} to interpolate in the Kurucz and 
MARCS APOGEE grids, as well as in the Phoenix grid mentioned above.

Finally, a 
\href{https://github.com/callendeprieto/mkk-atlas9}{convenient Perl script} to compute Kurucz model atmospheres is available using 
Luca Sbordone's port of ATLAS9 
\citep{2007IAUS..239...71S}. 
The port of ATLAS9 with updated molecular data and reference 
solar abundances used in the Kurucz APOGEE grid \citep{2012AJ....144..120M} can be obtained from 
a request to the authors. 

\section{Spectral synthesis}

There are truly dozens of packages for computing synthetic spectra from
model atmospheres. Limiting ourselves to open software we can mention 
the very popular 
\href{https://www.as.utexas.edu/~chris/moog.html}{MOOG} by \cite{1973PhDT.......180S}, 
\href{https://github.com/bertrandplez/Turbospectrum2019}{Turbospectrum} by \cite{2012ascl.soft05004P},
\href{https://www.appstate.edu/~grayro/spectrum/spectrum.html}{SPECTRUM} by R. Gray \citep{1994AJ....107..742G}, 
\href{https://kurucz.harvard.edu}{SYNTHE} by \cite{1993sssp.book.....K}, 
\href{https://tlusty.oca.eu/Synspec49/synspec.html}{Synspec} by \cite{2021arXiv210402829H}, 
or \href{https://github.com/JeffValenti/SME}{SME} by \cite{2017A&A...597A..16P}.
All these codes are written in FORTRAN, with the exception of SME, which is IDL,
and SPECTRUM, written in C.
In addition, \href{https://www.blancocuaresma.com/s/iSpec}{iSpec} by \cite{2014A&A...569A.111B} and \href{https://github.com/callendeprieto/synple}{Synple} by \cite{2021arXiv210402829H} offer powerful Python wrappers to one or several of the codes above. 

Synple is written in Python3 and gives a minimal, command-line interface. It wraps Synspec, and works with MARCS, Kurucz, Phoenix, and TLUSTY models. 
It requires only an input model atmosphere and the desired wavelength range. 
It attempts to make {\it clever} decisions for the user, for example regarding
which line lists to use or the frequency sampling. 

Synple includes
parallelization tools for using multiple cores on a single processor, or 
taking advantage of distributed machines using \href{https://slurm.schedmd.com/documentation.html}{Slurm}. It is particularly useful for building grids of 
models, streamlining the process (see Section \ref{grids}). 

\section{Comparing data to models}

I will emphasize some of the most useful tools available, focusing on 
those which are open source. We have mentioned SME and iSpec in the previous section for
spectral synthesis, and both include advanced tools for spectroscopic analysis.
In addition, \href{https://github.com/callendeprieto/FERRE}{FERRE} by 
\cite{2006ApJ...636..804A} is particularly well-suited for processing massive data sets. A Python
alternative, \href{https://github.com/callendeprieto/Synple}{BAS}, embedded in 
Synple, is now available. Other alternatives are MATISSE and GAUGIN \citep{2006MNRAS.370..141R,2016A&A...595A..18G}, \href{https://github.com/astroChasqui/q2}{q2} \citep{2014A&A...572A..48R}, or BACCHUS \citep{2016ascl.soft05004M}.

\subsection{FERRE}

FERRE is written in modern FORTRAN and parallelized with OpenMP. It is 
intrinsically multi-dimensional, with no constrains on the number of 
dimensions other than the practical limits on computing time, memory and 
numerical errors. It is  flexible to fit one, several or all the parameters
of the models adopted. FERRE employs pre-computed grids of synthetic 
spectra stored in RAM (or a database), which can be PCA-compressed. 

The speed of FERRE comes mainly from the fact that the synthetic spectra
are pre-computed. The code interpolates in the grids with linear, quadratic
or cubic polynomials. It has been successfully used in APOGEE, as well as 
in SDSS, DESI, WEAVE and other projects. FERRE includes several optimization 
algorithms such as the Nelder-Mead method \citep{nma}, the Boender-Timmer-Rinnoy Kan algorithm \citep{brk}, UOBYQA \citep{uobyqa}, or MCMC with differential evolution \citep{mcmc}. 

\begin{figure}
  \includegraphics[scale=0.8,angle=0]{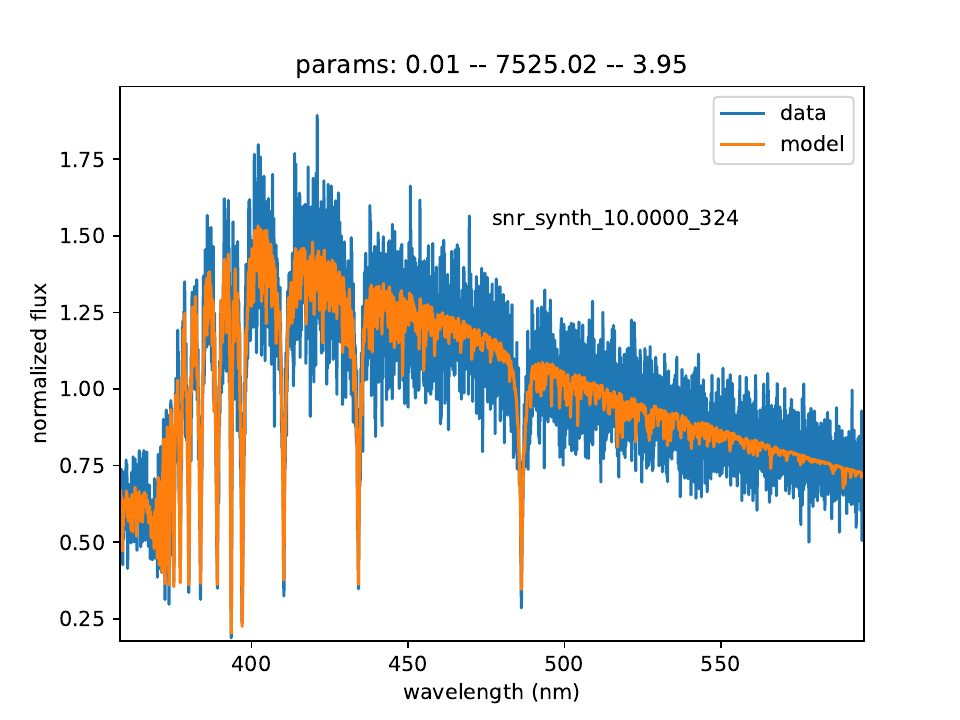}
  \includegraphics[scale=0.8,angle=0]{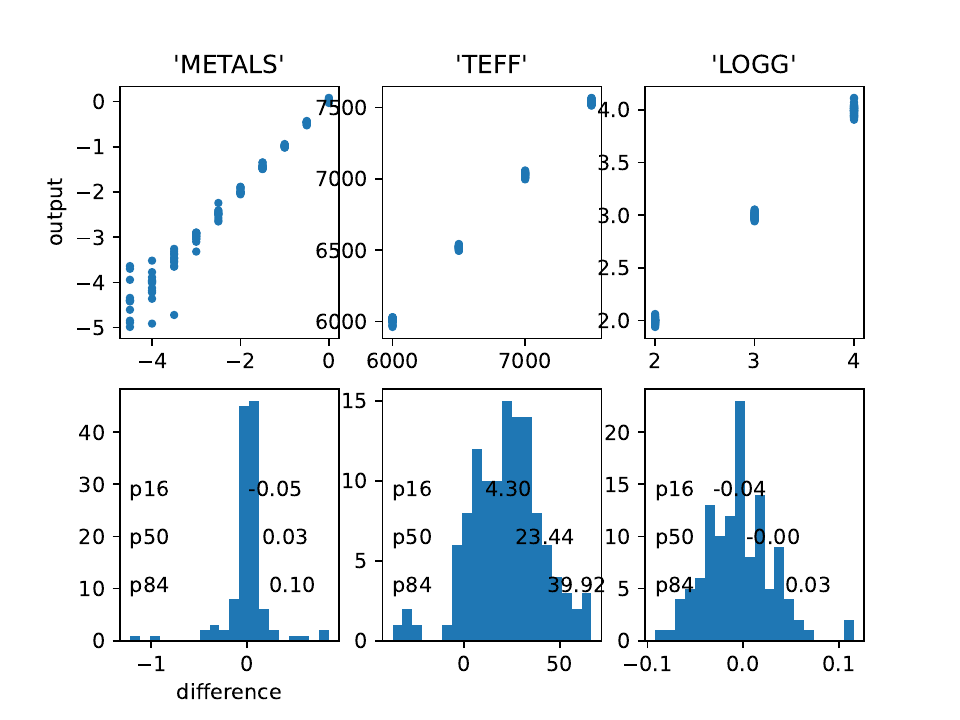}
  \caption{Comparison between the spectra (top panel, just for one case), and the parameters (bottom panels, 120 realizations) for simulated observations with a resolving power of about $R\sim 2000$ and a signal-to-noise  ratio  of 10. These data correspond to the example described in the FERRE manual.}
  \label{f2}
\end{figure}

Figure \ref{f2} shows an example of fitting (simulated) medium-resolution 
spectra with FERRE using only three free parameters: $T_{\rm eff}$, $\log g$ and [Fe/H]. Local algorithms such as Nelder-Mead or UOBYQA are usually the fastest, while MCMC is most useful when there are multiple local minima and accurate estimates of the uncertainties  are required. 

\begin{figure}
  \includegraphics[scale=0.5,angle=-90]{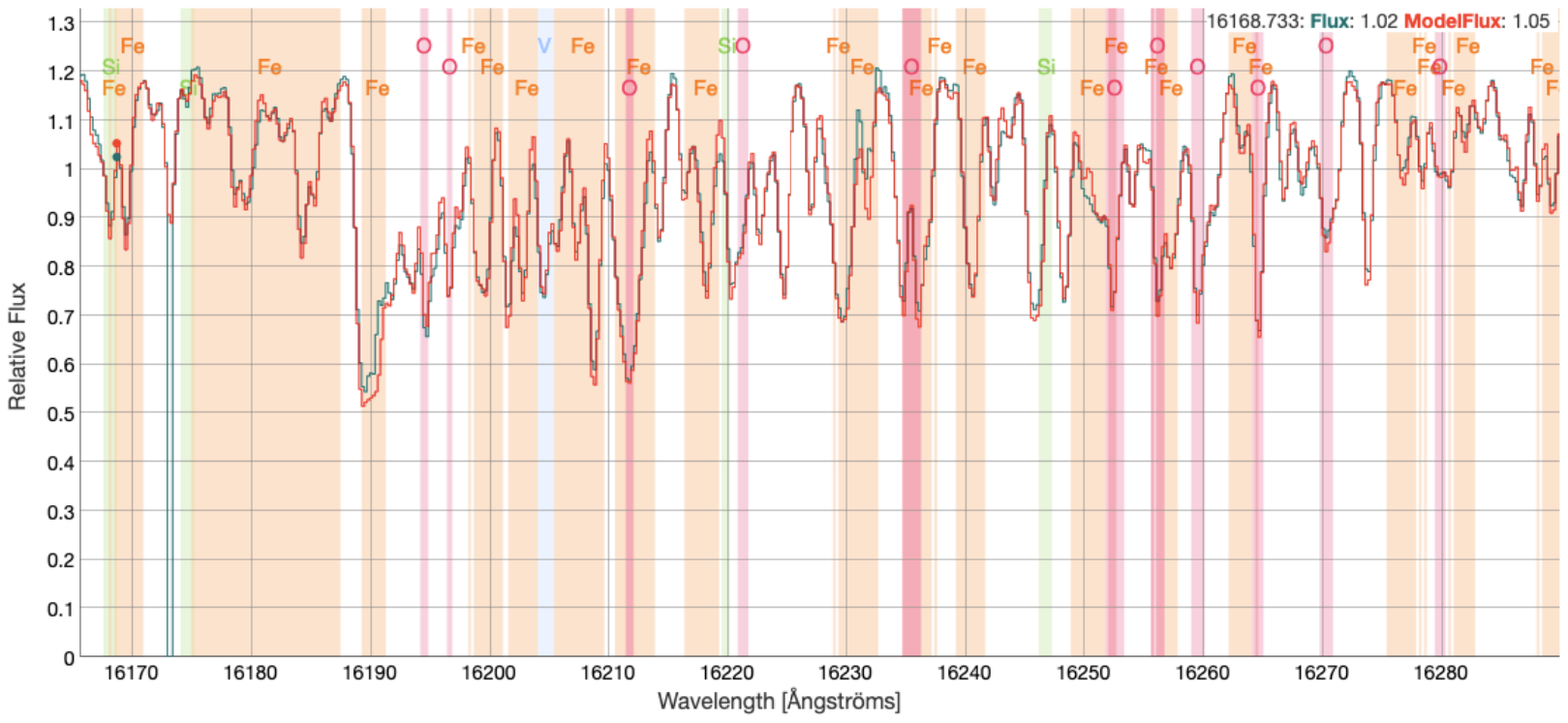}
  \caption{Comparison between an observed spectrum (green) and the best fit found with FERRE (red) for a particular  star included in the APOGEE data base. Some of the spectral features used to determine abundances of various elements (O from OH, Si from Si I lines or Fe from Fe I, for example) are shown. Plot made with the APOGEE webapp at the \href{https://dr17.sdss.org/infrared/spectrum/view/stars}{DR17  website}.}
  \label{f3}
\end{figure}

FERRE has been used to process in an automated fashion all the APOGEE 
high-resolution ($R\sim 22,500$) $H-$band 
spectra for about 3/4 of a million stars, 
and deriving abundances for over a dozen elements. In APOGEE, a first pass
fitting the entire 1500-1700 nm spectral range is made to derive simultaneously
$T_{\rm  eff}$, $\log g$, [Fe/H], [$\alpha$/Fe], [C/Fe], [N/Fe], and $v \sin i$
for dwarfs. A second pass is made, fitting one other element at a time by varying the overall metallicity but evaluating the $\chi-$square in the regions dominated by line absorption produced by the element of interest. A typical fitting for a small section of an APOGEE spectrum is shown in Fig. \ref{f3}.

FERRE is a bit too heavy to deploy for one or few stars.  Radial velocities 
have to be measured beforehand with other tools to place the observed 
spectra on the rest frame. 
Fitting [Fe/H] to derive abundances as done in APOGEE 
does not generally work at lower resolution. Another issue is that 
5-7D fits, which ideal for high-resolution spectra,
 require multiple starting points for local algorithms, 
and global ones (MCMC) are quite expensive in terms of computing time.
Regular grids, where the range and step for a given 
parameter are the same regardless 
of the values of the other parameters,  
are required by FERRE, and this means increased 
complexity for handling large ranges in parameters --
size limitations are already hit for APOGEE with $10^4$ frequencies, 
7 parameters, and  grids with millions of models.
In addition, gridding effects tend to appear with large
numbers of dimensions. When one or few spectra need to be analyzed automatically, BAS, described below, may pose a more practical alternative.

\subsection{BAS}

Interpolation of the models as done in FERRE reduces the time it takes 
model evaluation, roughly speaking from seconds to milliseconds,
 but the speed bottleneck is usually accessing RAM.
Pre-computing the interpolations can save additional time, and that is the 
main idea beyond BAS, the Bayesian Algorithm in Synple.
In BAS the use of a Bayesian scheme brings a global algorithm, 
and more reliable error bar estimates. Another nice feature is that BAS
is pure Python, and is integrated in Synple.

\begin{figure}
  \includegraphics[scale=0.5,angle=-90]{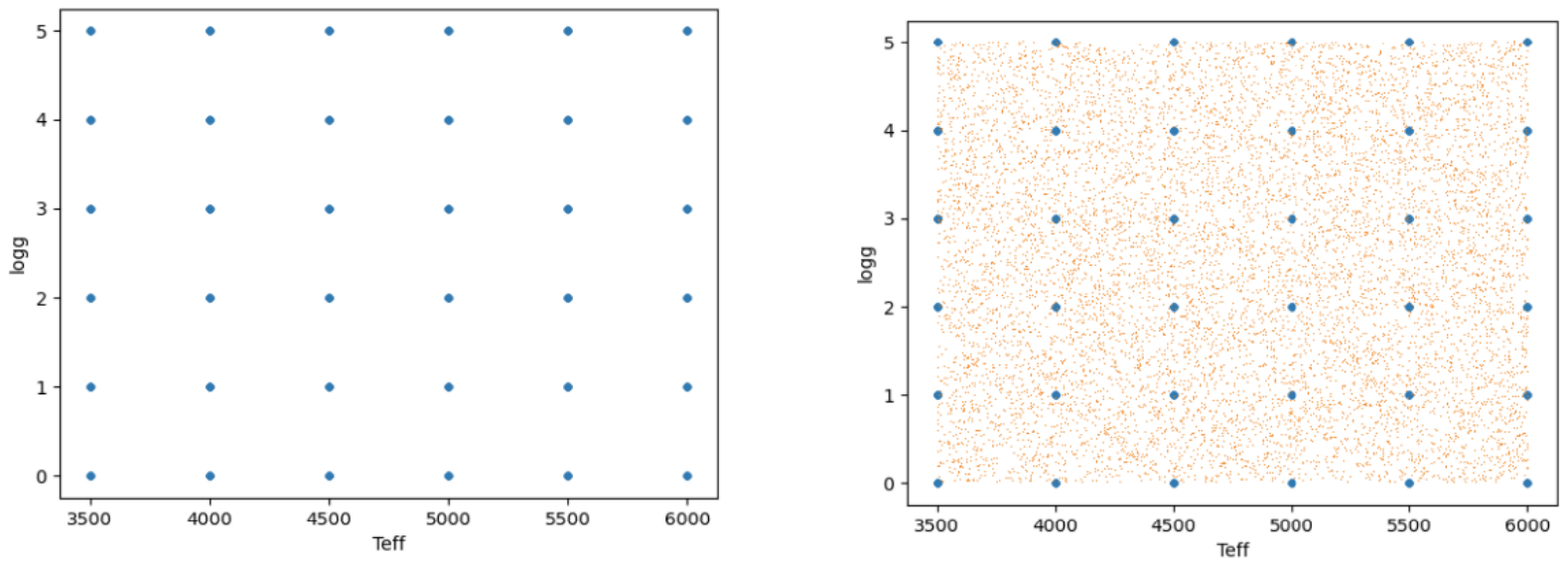}
  \caption{Parameters for a regular 2D model ($T_{\rm eff}$, $\log g$) grid 
such as those used in FERRE (left-hand panel), compared to an irregular grid (red points on the right panel) as used in BAS.}
  \label{f4}
\end{figure}

BAS adopts the simplest possible Bayesian scheme: it uses flat priors and the
input grid samples the full model space randomly. The expectation value for a parameter $x$ is computed by summing over all the models with parameters  $x_i$ and 
$\chi-$square $\chi_i$
\begin{equation}
E(x) \propto \sum_i x_i e^{-\chi_i/2},
\end{equation}
and its covariance with a parameter $y$ is
\begin{equation}
Covar(x,y) \propto \sum_i (x_i-E(x))  (y_i-E(y))  e^{-\chi_i/2}.
\end{equation}

BAS uses all the models in an interpolated irregular grid as illustrated in \ref{f4}: it 
may do a lot more work evaluating the merit function than an
 optimization algorithm, but the number of evaluations is 
always the same, which brings new opportunities for parallelization,
 in particular with modern graphics cards.

\begin{figure}
  \includegraphics[scale=0.5,angle=-90]{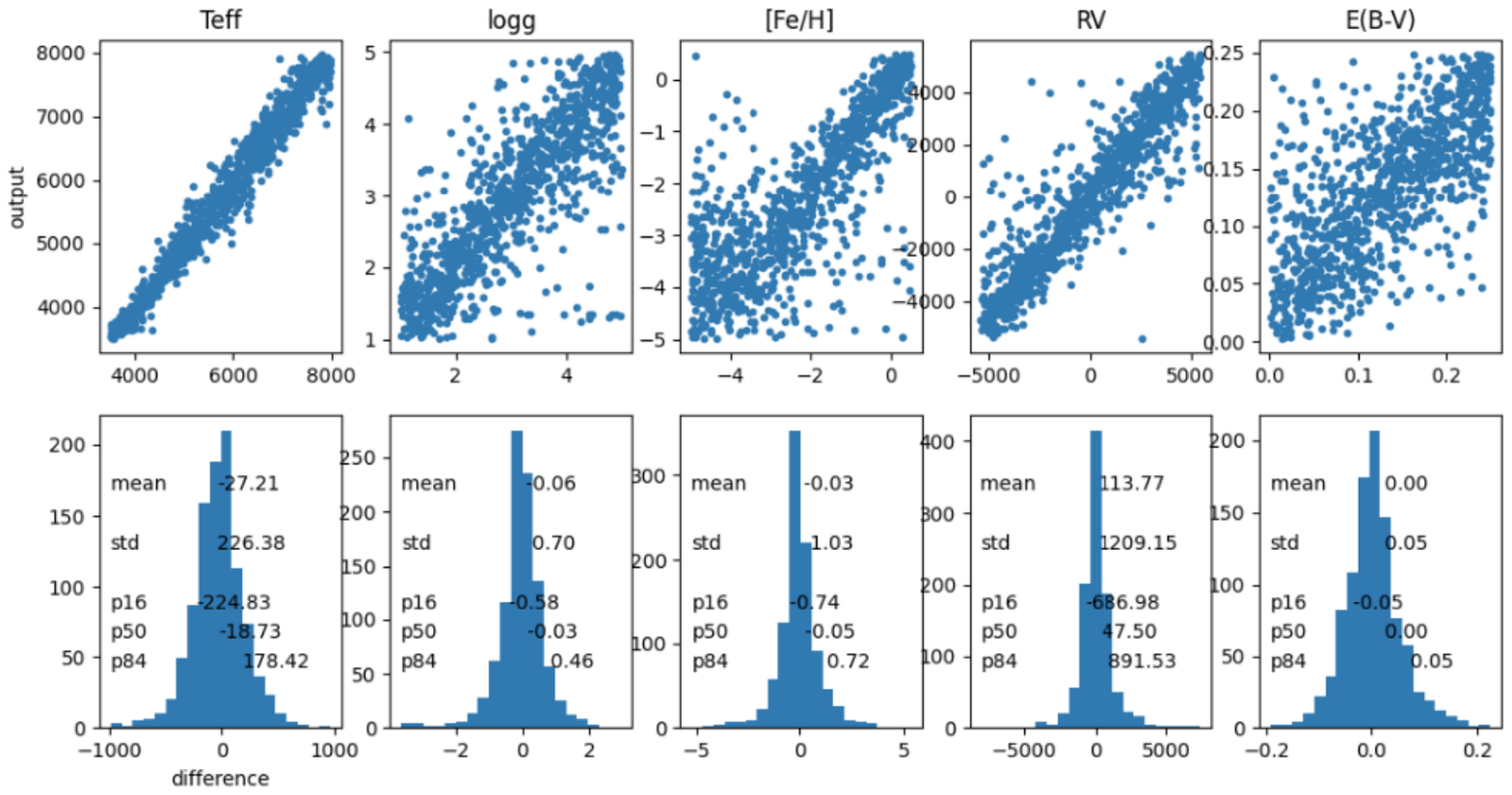}
  \caption{Output from the BAS self-evaluation routine for a grid with Gaia Bp/Rp spectra spanning $3500 < T_{\rm eff} < 8000$ K. The analysis includes the simultaneous determination of $T_{\rm eff}$, $\log g$, [Fe/H], radial velocity and interstellar reddening. Similar to the bottom panels in Fig. \ref{f2}, the upper  panels compare input (ground-truth) and output parameters, while the bottom panels show the distribution of residuals.} 
  \label{f5}
\end{figure}

BAS implements a number of features that bring it one step closer to
full automation. For example,  
BAS recognizes the source of the data by inspecting the input files' content, 
and them uses an appropriate model grid.
There are already grids available for K-F-G-A type stars and white dwarfs,
customized 
for INT-IDS (MILES), HST-STIS/NICMOS, GTC-OSIRIS, Mayall-DESI, Gaia XP, or 
LAMOST data. 
BAS also incorporates cross-correlation and template matching for 
RV determination, and can apply   
on-the-fly reddening corrections. In addition, BAS includes a self-evaluation
module, illustrated in Fig. \ref{f5}, which takes an input grid, creates new templates using radial-basis-functions (RBF) interpolation, adds noise and analyzes the simulated spectra to
evaluate performance. 

\section{Grid building}
\label{grids}

Synple comes with a suite of tools to facilitate the task of building grids of synthetic spectra suitable for FERRE, BAS or other applications. Among these utensils, we can mention the functions {\tt polysyn}, for organizing parallel computations for an arbitrary
number of input models, {\tt synth\_rbf} for RBF interpolation, or {\tt create\_irregular\_grids} for setting up calculations of Kurucz model atmospheres using ATLAS9.
There are tools available as well to read and merge grids. 

\section{Advanced modeling}

The framework discussed so far is limited to classical model atmospheres: 
one-dimensional models in hydrostatic balance and assuming local thermodynamical equilibrium (LTE). But precision work requires refinements such as considering departures from LTE and hydrostatic equilibrium.

The overall strategy of using pre-computed spectral grids works the same 
independently from the nature of the models, no matter how simple or complex they are.
Departures from LTE have already been implemented in the SDSS DR7 APOGEE grids 
(Na, Mg, K, Ca) or in the SME analysis in GALAH.
Note that NLTE effects can propagate from one element to another 
\citep{2020A&A...637A..80O}, making this business more complicated.

Three-dimensional hydro dynamical models are scarce, and therefore 
grid density is an issue, but it is possible to use a grid of flux 
ratios F(3D)/F(1D) on which interpolation can be employed to correct 
high-density F(1D) grids \citep{2022A&A...661A..76B}.
Using 3D models requires doing NLTE in 3D, and we emphasize that
taking the horizontal average of a 3D model gives us a 1D model, not a 3D one. 

\section{Closing conclusions}

Full automation means passing a reduced spectrum and getting all possible information back (parameters and abundances).
We do not have such software yet,
but we need it badly. 
It will likely be available within 1-2 years, and we already have decent
 approximations for some data sets.
Progress is hindered by keeping codes and models to ourselves, so please be open and share them!

While the path to perform fully automated analyses with one-dimensional (LTE or NLTE) models is fairly well defined, this is not yet the case for three-dimensional (radiation-hydro dynamical) models, where limited model availability brings
additional difficulties.

{\bf Acknowledgedments}

I thank the organizers for their kind invitation and congratulate them for an excellent job. Congratulations, Beatriz, for your unique  and many contributions to science! 

I am grateful for financial support from the Spanish government (grants AYA2017-86389-P and PID2020-117493GB-I00). This research made use of computing time available on the high-performance computing systems at the Instituto de Astrofisica de Canarias, part of the Spanish Supercomputing Network, and at the US National Energy Research Scientific Computing Center.

\end{document}